\documentclass[preprint]{aastex631}

\begin{document}

\title{The NEO Surveyor Near Earth Asteroid Known Object Model}

\correspondingauthor{Tommy Grav}
\email{tgrav@arizona.edu}

\author[0000-0002-3379-0534]{Tommy Grav}
\affiliation{University of Arizona\\
Lunar and Planetary Laboratory \\
1629 E University Blvd\\
Tucson, AZ 85721-0092, USA}

\author[0000-0002-7578-3885]{Amy K. Mainzer}
\affiliation{University of Arizona,Lunar and Planetary Laboratory, 1629 E University Blvd, Tucson, AZ 85721-0092, USA}

\author[0000-0003-2638-720X]{Joseph R. Masiero}
\affiliation{IPAC, California Institute of Technology, 1200 E. California
             Blvd, Pasadena, CA 91125, USA}

\author[0000-0003-1876-9988]{Dar W. Dahlen}
\affiliation{IPAC, California Institute of Technology, 1200 E. California
             Blvd, Pasadena, CA 91125, USA}

\author{Tim Spahr}
\affiliation{NEO Sciences}

\author[0000-0002-1804-7814]{William F. Bottke}
\affiliation{Southwest Research Institute, Solar System Science and Exploration Division, 1050 Walnut St, Suite 300, Boulder, CO 80302}

\author[0000-0002-8532-9395]{Frank J. Masci}
\affiliation{IPAC, California Institute of Technology, 1200 E. California
             Blvd, Pasadena, CA 91125, USA}



\begin{abstract}

The known near-Earth object (NEO) population consists of over 32,000 objects, with a yearly discovery rate of over 3000 NEOs per year. An essential component of the next generation of NEO surveys is an understanding of the population of known objects, including an accounting of the discovery rate per year as a function of size. Using a near-Earth asteroid (NEA) reference model developed for NASA's NEO Surveyor (NEOS) mission and a model of the major current and historical ground-based surveys, an estimate of the current NEA survey completeness as a function of size and absolute magnitude has been determined (termed the Known Object Model; KOM). This allows for understanding of the intersection of the known catalog of NEAs and the objects expected to be observed by NEOS. The current NEA population is found to be $\sim38\%$ complete for objects larger than 140m, consistent with estimates by \citet{Harris.2021a}. NEOS is expected to catalog more than two thirds of the NEAs larger than 140m, resulting in $\sim76\%$ of NEAs cataloged at the end of its 5 year nominal survey \citep{Mainzer.2023}, making significant progress towards the US Congressional mandate. The KOM estimates that $\sim77\%$ of the currently cataloged objects will be detected by NEOS, with those not detected contributing $\sim9\%$ to the final completeness at the end its 5 year mission. This model allows for placing the NEO Surveyor mission in the context of current surveys to more completely assess the progress toward the goal of cataloging the population of hazardous asteroids.

\end{abstract}



%% ========================= SECTION ======================================
\section{Introduction} 
\label{sec:intro}

The near-Earth object (NEO) population is made up of asteroids and comets that are on orbits that take them close to Earth. NEOs are defined as objects with orbits that bring them closer than 1.3 au (perihelion distance $q \le 1.3$ au) from the Sun. They consist of both active and non-active bodies, where the near-Earth comets make up about $5-15\%$ of the NEO population \citep{Wetherill.1987a,Wetherill.1988a,Bottke.2002a,Bauer.2017a,Granvik.2018}. In this paper we will focus on the Near-Earth Asteroid (NEA) population exclusively.

\begin{figure}[ht!]
\plotone{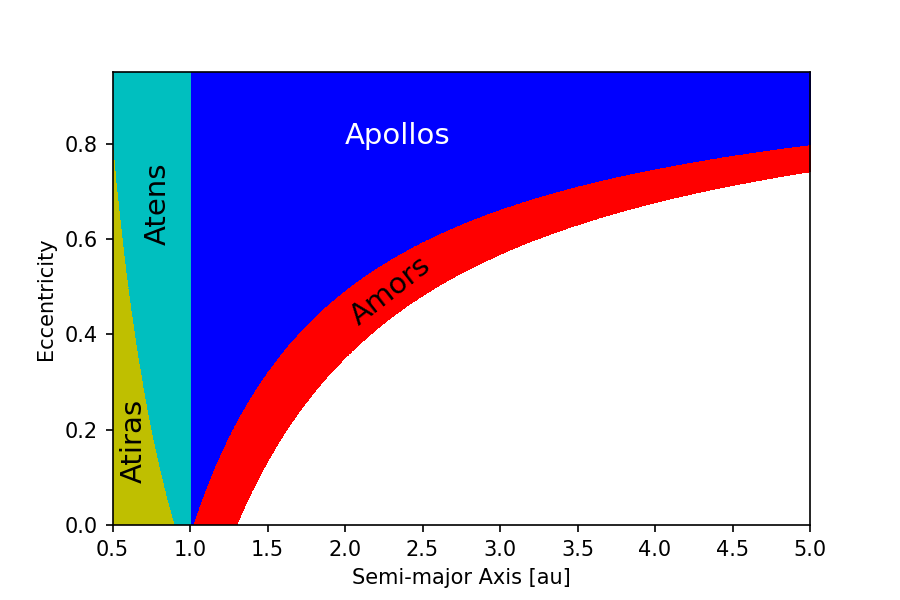}
\caption{A schematic showing the region in semi-major axis vs. eccentricity space of the four NEO subpopulations. 
\label{fig:subpop_chart}}
\end{figure}

The NEAs can be split into four main sub-populations. The largest of these sub-populations is the Apollos, named after one of its members, (1862) Apollo. They are defined as objects with semi-major axis $a > 1.$ au, and perihelion distance $q \le 1.017$au, consisting mainly of objects whose orbits cross the orbit of the Earth. The Amors, named after its archetype object (1221) Amor, are defined as objects with perihelion distance $1.017 < q \le 1.3$ au. The Amors are the second largest group, consisting of objects whose orbits are entirely outside the Earth's orbit. The Atens are defined as objects with semi-major axis $a < 1$ au and aphelion distance $Q > 0.983$ au. They consist of objects with orbits that are mainly inside the orbit of the Earth, but cross the orbit of the Earth. The smallest of the four sub-populations are the Atiras, named after the first known object of its kind (163693) Atira. These objects have orbits with aphelion $Q \le 0.983$ au, putting their entire orbit inside the Earth's orbit. Figure \ref{fig:subpop_chart} shows the four sub-populations in the semi-major axis and eccentricity space. 

%% Figure 1

The Minor Planet Center (MPC) \footnote{https://minorplanetcenter.net} maintains the official catalog of observations and orbital elements for asteroids, comets and natural satellites in our solar system. At the beginning of June 2023, the MPC Catalog contained orbits and observations of more than $32,100$ NEAs, of which about a quarter are Potentially Hazardous Asteroids (PHAs), which have Minimum Orbital Intersection Distance (MOID) of $\le 0.05$ au. Usually the definition of PHAs include a selection criterion of $H_V \le 22$ mag, with this being traditional used as a proxy for a diameter of $\ge140$ m. However, in this work we consider any object with MOID $\le 0.05$ au to be a PHA, as a dark object with diameter of 140m can have absolute magnitude as faint as $H_V \sim 24$ mag. 

%% Figure 2
\begin{figure}[ht!]
\plotone{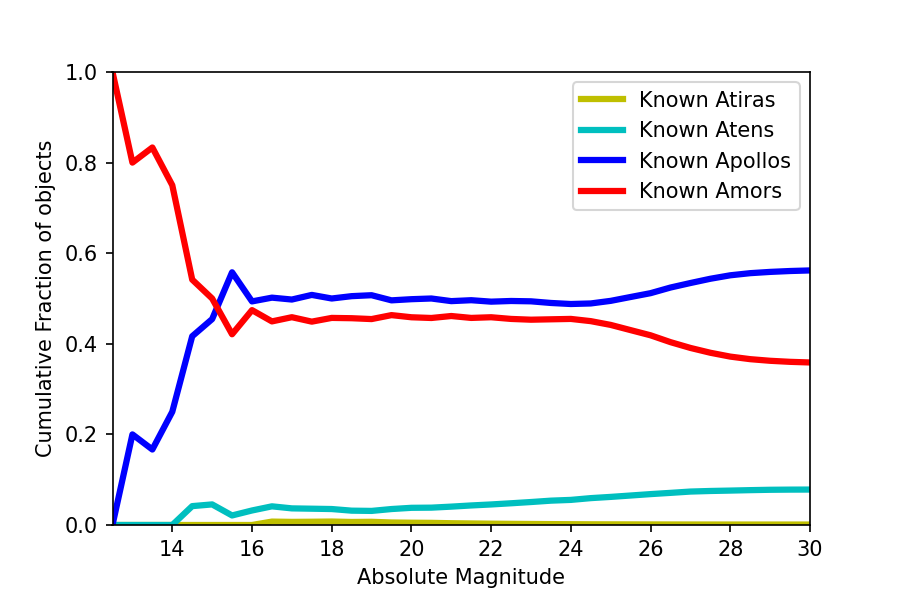}
\caption{The fraction of each NEA sub-population (Atiras, Atens, Apollos, and Amors) from the MPC catalog as a function of absolute magnitude. The fractions for a value of absolute magnitude are cumulative, including all objects with values less than the value plotted. The effects of observational bias are clearly seen at $H_V > 24$ magnitude.
\label{fig:subpop_fraction}}
\end{figure}

 The Apollos make up $\sim51.2\%$ of this cataloged dataset; $\sim40.8\%$ of the cataloged NEAs are Amors and $\sim7.8\%$ are Atens. Less than $0.2\%$ of the known NEAs are Atiras.  Figure \ref{fig:subpop_fraction} shows that the cumulative fractions are relatively stable up to about absolute magnitude $H_V \sim 20$ mag, but for higher absolute magnitudes the fraction of NEAs that are Atens starts rising slightly. The relative fraction of Amors and Apollos remain stable until $H_V\sim 24$ mag; at higher absolute magnitude values, the fraction of Apollos starts rising significantly. These fractions are influenced by observational biases, especially at smaller sizes (higher absolute magnitudes), due to the observational geometry. A closer look at these observational biases will be discussed in Section \ref{sec:kom}. 

In studies of NEAs, two diameter regimes have become important, the objects thought to be larger than 1km, and the population between 140m and 1km. The largest objects, with effective diameters larger than 1km, impact infrequently. But these objects possess the power to cause global extinction events \citep{Alvarez.1980a}. It is estimated that these objects posed about $90\%$ of the risk from asteroid impacts, and the ``SpaceGuard" \citep{Milani.1990a,Morrison.1992a} goal of discovering $90\%$ of asteroids in this size range was created to address this risk. This goal was found to have been completed around 2010 \citep{Mainzer.2011a,Granvik.2018}. Community studies \citep{Stokes.2003a,Stokes.2017a,NAP25476} have found that after the Spaceguard goal was achieved, the majority of remaining risk lay with the asteroids larger than $140$m, which are capable of creating significant local or regional damage upon impact. The George E. Brown, Jr. Near-Earth Object Survey Act was passed by the United States Congress in 2005, directing that NASA by 2020 detect and track more than $90\%$ of all NEOs larger than 140 m in effective diameter\footnote{https://www.congress.gov/bill/109th-congress/house-bill/1022/text}. New discoveries have been dominated by the Catalina Space Survey (MPC observatory codes 703 and G96) and Pan-STARRS (MPC observatory codes F51 and F52) over the last decade, together accounting for $80-92\%$ of discoveries every year covering that time period. In section \ref{sec:discrate} we will take a closer look at the discovery statistics over the last decade and their implications for understanding the current completeness level of the known NEA population. 

One of the fundamental problems in the field of planetary defense is that the goals outlined by Spaceguard and the George E. Brown, Jr. Act are defined in diameter, since impact energy depends on diameter to the third power ($\sim D^3$); therefore accurate knowledge of diameter is important for constraining impact energy. The current major surveys obtain observations at visible wavelengths, allowing for the derivation of orbital parameters and absolute magnitude, but these bandpasses do not allow for direct derivation of the effective spherical diameter. The absolute magnitude at visible wavelengths, $H_V$, is defined as the theoretical visible magnitude an object would have if it was 1 au from the Sun and Earth at zero phase angle. If the reflectivity of the object is known, i.e. the visible geometric albedo, the size of the object can be determined by $D = (1329/\sqrt{p_V}) * 10^{-0.2H}$ \citep{Fowler.1992a}, where $D$ is the diameter in kilometers, $p_V$ is the geometric albedo at visible wavelengths, and $H_V$ is the absolute magnitude. Traditionally an albedo of $p_V = 0.14$ has been assumed, which yields $H_V = 22$ mag for an object with a diameter of 140 meters and $H_V = 17.75$ mag for an object with a 1 kilometer diameter. However, infrared missions like IRAS \citep{Tedesco.1992a,Tedesco.2002a}, NEOWISE \citep{Mainzer.2011a,Mainzer.2011e,Wright.2016a} and Akari \citep{Usui.2011a} showed that the geometric albedo for asteroids varies from about $2\%$ to $60\%$, with the dark component of the population having a Maxwellian distribution peaking at $\sim3\%$ and the rest in a Maxwellian distribution peaking at $\sim17\%$ \citep{Wright.2016a}. About $\sim40\%$ of the NEOs of a given size have $p_V < 0.1$ \citep{Mainzer.2011b}. This introduces complexity in understanding the completeness for effective diameter-limited samples when a majority of observations are obtained at visible wavelengths. \cite{Wright.2016a} examined what equivalent $H_V$ magnitude would be required in order to reach $90\%$ completeness for NEAs larger than 140m considering the double Maxwellian distribution of the NEA albedo distribution derived by the NEOWISE mission. They found that in order to reach $90\%$ survey completeness for this size range, a $90\%$ completeness of objects with $H_V \le 23$ mag is needed. We thus use $H_V  \leq 23$ mag as the appropriate proxy for objects larger than 140m in this paper, rather than the traditional $H_V = 22$ mag. 

In this paper we examine the known population of NEAs and how this population will be observed by the NEO Surveyor mission \citep[NEOS;][]{Mainzer.2023} and will compliment its expected performance. NEOS is part of the the next generation surveys for asteroids that could impact the Earth. Previously known as NEOCam \citep{Mainzer.2015a}, NEO Surveyor is a space mission designed to detect, track and characterize NEAs using thermal emission observations. It recently passed NASA's preliminary design review phase and is scheduled to launch in September 2027. It is designed to catalog more than two thirds of the PHAs larger than 140m in diameter by the end of its nominal 5 year mission. NEO Surveyor uses a reference model of synthetic asteroids and comets to gauge its performance and progress against this goal. See \citet{Mainzer.2023} for additional information on the NEOS reference model. However, since the catalog of NEAs and PHAs is not empty at the time of the start of the NEO Surveyor mission, it is important to determine a set of proxy objects in the synthetic reference model that represents the currently known objects. This allows for understanding of what types of objects among the current MPC catalog are unlikely to be detected by NEO Surveyor and thus would count towards the cataloging goal, and which objects that are currently known are likely to be also be detected by NEO Surveyor, providing both optical and thermal observations that allow for determination of both diameter and geometric albedo. While future ground based surveys, such as the Vera Rubin Observatory \citep{Jones.2018a,Veres.2017a} will also contribute to the total number of cataloged NEOs, the actual performance of these future surveys are still in the planning phases, which could potentially have large impacts on their performance in terms of NEO discovery. A study of the performance of the Vera Rubin Observatory (then the Large Synoptic Sky Telescope; LSST) and its synergy and overlap with NEO Surveyor (then the NEOCam mission) can be found in \citet{Grav.2016a}. Thus this paper provides a worst case scenario, where no additional future NEAs are assumed to have been discovered prior to the NEO Surveyor launch.

We note that this paper uses data from the MPC catalog extracted on Sep 05, 2023. The orbits and absolute magnitudes of objects cataloged by the MPC is in a constant state of flux as additional observations are continuously being submitted by observers from around the world. In this paper we compare our results to that of \citet{Harris.2021a}, but we note here that recent re-calculations in the absolute magnitudes for NEOS cataloged for the largest has yielded a revision in number of large NEOs (Harris, personal communications)\footnote{https://www.hou.usra.edu/meetings/acm2023/pdf/2519.pdf}. The revision resulted in a change of $\sim100$ less NEOs with $H_V < 17.75$ mag compared to the data used in \citet{Harris.2021a}, meaning that the NEOs were on average revised $0.13$ magnitudes fainter. This is in line with \citet{Pravec.2012a}, comparing the absolute magnitudes from the MPC catalog with a list of asteroids with high precision photometric observations, which found that the absolute magnitudes in the MPC catalog were on average $0.3$ magnitude too bright.
 
 We examine the current status of the NEA and PHA populations cataloged by the Minor Planet Center (section \ref{sec:known_nea_pop}). In section \ref{sec:NEOS} we discuss the NEO Surveyor mission, along with the NEOS Survey Simulator (NSS) and the NEOS reference model, two tools used to predict the performance of the NEO Surveyor mission in detecting, tracking, and characterizing small bodies in our solar system during its 5-year nominal mission. The NEOS Known Object Model is described in section \ref{sec:kom}; this model is applied to the NEOS reference model to determine which of the synthetic objects would be expected to be present in the currently known population. Section \ref{sec:NEOSandKOM} examines what portion of the known population is detected by the NEOS survey and which objects remain undetected at the end of the 5-year nominal mission. Finally, section \ref{sec:conclusions} discusses the findings of this paper.

%% ========================= SECTION ======================================
\section{The Known NEA Population}
\label{sec:known_nea_pop}

At the beginning of June 2023, the MPC catalog contained more than $32,100$ NEAs. Of these, just over 3000 NEAs have been numbered, indicating that their orbits are well known and do not require additional observations to maintain accurate positional predictions over the next century. An additional $\sim2000$ NEAs have observations spanning more than $10$ years, with $10\%$ of these spanning more than $20$ years, all of which should become numbered in the near future. Only a third of the known NEAs have multi-opposition orbits, with observations at two or more opposition epochs. A majority of the NEAs are only observed during their discovery apparitions, with almost half having been observed for less than 7 days, leaving them basically lost and in need of re-discovery to further refine their orbital parameters. 

 As mentioned above, the NEA population is divided into four sub-populations. However, the MPC catalog has a heavy observational bias towards certain of these sub-populations, especially the Atens and Apollos, as these objects tend to come much closer to the Earth and can thus be observed at much smaller sizes. Therefore, it is important to look at absolute magnitude limited samples of the NEAs when considering fractions. When looking at the population of $853$ objects with $H_V < 17.75$ magnitude, the fraction of objects in the sub-populations are $0.9\%$, $3.9\%$, $50.1\%$, and $45.1\%$ for the Atiras, Atens, Apollos, and Amors, respectively. For the more than $13,000$ NEAs with  $H_V < 23$ magnitude, the proxy we are using for 140m, the fractions have changed slightly to $0.2\%$, $5.1\%$, $49.2\%$ and $45.5\%$ for the Atens, Apollos and Amors, respectively. 

Figure \ref{fig:subpop_fraction} shows the fractions for the three sub-populations when including all objects with a certain absolute magnitude range. The fractions stay relatively consistent to about $H_V \sim 24$ magnitude, where the observational effects favoring discovery of Apollos and Atens become apparent. There is an increase in the number of Atens from $3.9\%$ at $H_V \le 17.75$ magnitude to $5.1\%$ at $H_V \le 23$ magnitude, which comes at the expense of the Atira and Apollo sub-populations. This increase in numbers could be due to the effect of observational biases being more strongly apparent at smaller absolute absolute magnitudes than those of the Apollos and Amors. Alternatively, the change could represent a real increase in Atens compared to other populations due to a difference in the source populations replenishing this sub-population \citep{Bottke.2002a}. The relative fractions of objects discussed in this section are used when generating the NEO Surveyor reference model, which generates a synthetic population of NEAs by using physical parameter models derived using the NEOWISE mission for each sub-population \citep{Mainzer.2012b,Mainzer.2023}. 

\subsection{The Discovery Rate of Near-Earth Objects}
\label{sec:discrate}

\begin{table}[th]
    \centering
    \begin{tabular}{|l|r|rr|rr|rr|}
    \hline
    Year & Total & \multicolumn{2}{c|}{$H \leq 17.75$} & 
        \multicolumn{2}{c|}{$17.75 < H \leq 23$} &
        \multicolumn{2}{c|}{$23 \leq H$} \\
    \hline
     2011 &  897 & 19 & $2.1\%$ &  477 & $53.2\%$ &  401 & $44.7\%$ \\
     2012 &  991 & 15 & $1.5\%$ &  480 & $48.4\%$ &  496 & $50.1\%$ \\
     2013 & 1029 & 11 & $1.1\%$ &  500 & $48.6\%$ &  518 & $50.3\%$ \\
     2014 & 1480 &  8 & $0.5\%$ &  651 & $44.0\%$ &  821 & $55.5\%$ \\
     2015 & 1551 &  7 & $0.5\%$ &  688 & $44.4\%$ &  856 & $55.2\%$ \\
     2016 & 1874 &  7 & $0.4\%$ &  731 & $39.0\%$ & 1136 & $60.6\%$ \\
     2017 & 2039 &  7 & $0.3\%$ &  743 & $36.4\%$ & 1289 & $63.2\%$ \\
     2018 & 1825 &  5 & $0.3\%$ &  614 & $33.6\%$ & 1206 & $66.1\%$ \\
     2019 & 2438 &  6 & $0.3\%$ &  750 & $30.8\%$ & 1682 & $69.0\%$ \\
     2020 & 2958 &  3 & $0.1\%$ &  829 & $28.0\%$ & 2126 & $71.9\%$ \\
     2021 & 3093 &  5 & $0.2\%$ &  730 & $23.6\%$ & 2358 & $76.2\%$ \\
     2022 & 3189 &  4 & $0.1\%$ &  725 & $22.7\%$ & 2460 & $77.1\%$ \\
     \hline
    \end{tabular}
    \caption{The yearly discovery statistics over the last decade. Note that the H magnitude of all NEAs evolve as more observations are reported, thus the reported numbers above might vary by a few objects when using a different MPC catalog instance than used here. }
    \label{tab:disc_rate}
\end{table}

\begin{figure}[ht!]
\plotone{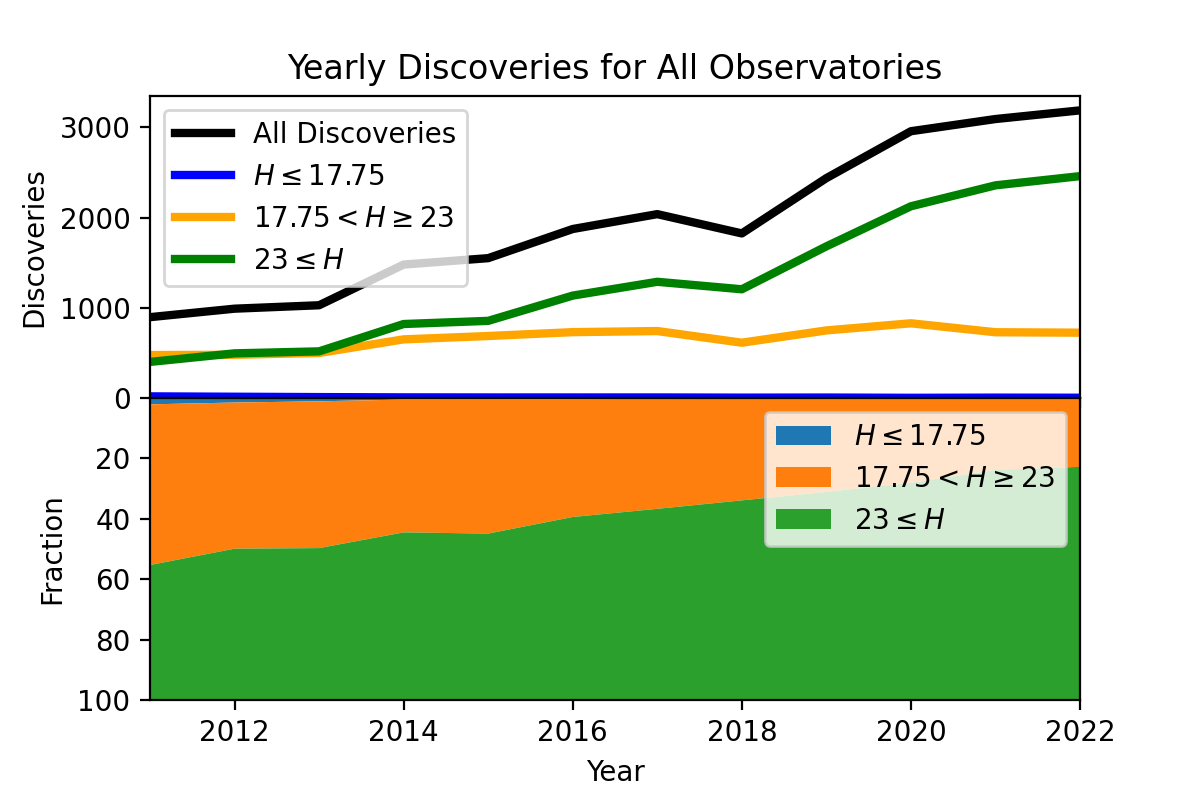}
\caption{The discovery statistics over the last decade show that while the number of discoveries per year has risen regularly year over year, the fraction of objects with $H_V \leq 23$ mag has dropped by half from over half of the discoveries in 2011 to under a quarter of the discoveries in 2022 . 
\label{fig:table1plot}}
\end{figure}

\begin{figure}[ht!]
\plotone{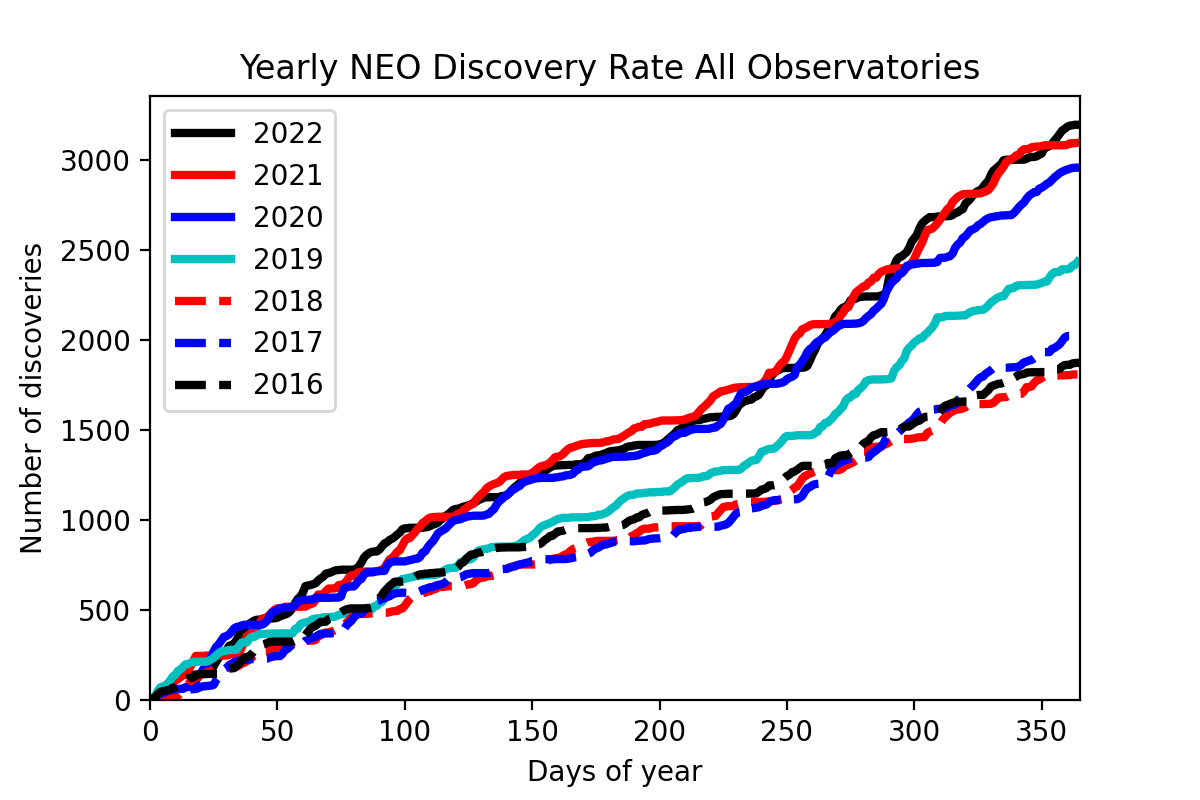}
\caption{The yearly NEA discovery rate from 2016 to 2022. 
\label{fig:disc_rate}}
\end{figure}

The number of discoveries of NEAs per year has sharply increased over the last few decades, from $27$ discoveries in 1992, $485$ discoveries in 2002, $991$ discoveries a decade ago, to $3,189$ discoveries in 2022 (see Table \ref{tab:disc_rate}, Figure \ref{fig:table1plot}, and Figure \ref{fig:disc_rate}). These steady increases are due to the increases in effort, funding and improvement in technology as touched upon in Section \ref{sec:intro}. While the total number of discoveries has sharply increased over the last decade, there has been a significant shift towards discovery of smaller objects. In 2012, objects with $H_V < 23$ mag made up almost $50\%$ of all discoveries, but in 2022 that number had declined to about $23\%$ of discoveries for that year. This trend mirrors the trend seen in the largest objects ($H_V \leq 17.75$ mag), which constituted more than half of discoveries in the mid-1980s. By the late-1990s they only made up a quarter of the discoveries, even though the number of discovered objects had gone up dramatically (from $\sim 10s$ of discoveries to more than $200$ objects). Since then the number of discoveries per year of objects with $H_V \le 17.75$ mag stayed at more than $45$ objects for 8 years, before declining steadily year by year to the handful of objects discovered per year currently. This is due to the fact that the objects in this size regime are nearing observational completeness, with the remaining objects being increasingly difficult to discover due to their orbital geometry. 

Currently the number of yearly discoveries with $H_V \le 23$ mag has stabilized at an average of $731$ objects per year. \cite{Harris.2021a} found that there are between $31,341$ objects with $H_V < 22.75$ mag and $47,577$ objects with $H_V < 23.25$ mag, so we assume from this that there are $\sim 39,500$ objects with $H_V < 23$ mag. There are about $13,400$ objects with $H_V \le 23$ mag in the current MPC catalog, which means that to reach $90\%$ completeness for $H_V \le 23$ mag as prescribed by \cite{Wright.2016a} in order to insure $90\%$ completeness for NEAs larger than 140m, the current surveys would need to discover an additional $22,000$ objects with $H_V \le 23$ mag. At the current discovery rate this would take more than 30 years. As discussed for the objects with $H_V \leq 17.75$ mag, it is expected that the number of discoveries per year will decline as the completeness of the objects with $H_V \leq 23$ mag increases. This would significantly extend the time it would take for the current surveys to reach $90\%$ survey completeness for NEAs with diameters larger than 140 meter.

%% ========================= SECTION ======================================
\section{NEO Surveyor Mission}
\label{sec:NEOS}

The NEO Surveyor is a NASA mission designed to find, catalog, and characterize NEAs. It is a single- instrument 50 cm space telescope operating in two infrared wavelength channels, centered on $4.6\mu$m and $8\mu$m. At these wavelengths thermal emission from NEAs dominates the observed flux, due their surface temperatures of $200 - 300$K. Situated at the Sun-Earth L1 Lagrange point, the mission will perform a nominal 5 year survey, observing the field of regard of $45 - 120$ degrees solar ecliptic longitude angle in between $\pm 40$ degrees ecliptic latitude on either side of the sun. Each of these sides takes about $6 - 7$ days to complete, allowing NEO Surveyor to provide self-follow-up of its discoveries, as most of the moving objects will not have left the field of regard by the time the area is re-observed about two weeks later. NEO Surveyor is designed to catalog more than two thirds of the PHAs larger than 140m in diameter \citep{Mainzer.2023} by the end of its nominal 5 year mission, with current models showing that it will reach more than $80\%$ completeness for PHAs larger than $140$m. For a more in depth description of NEO Surveyor and its mission see \citet{Mainzer.2023}.

\begin{figure}[ht!]
\plotone{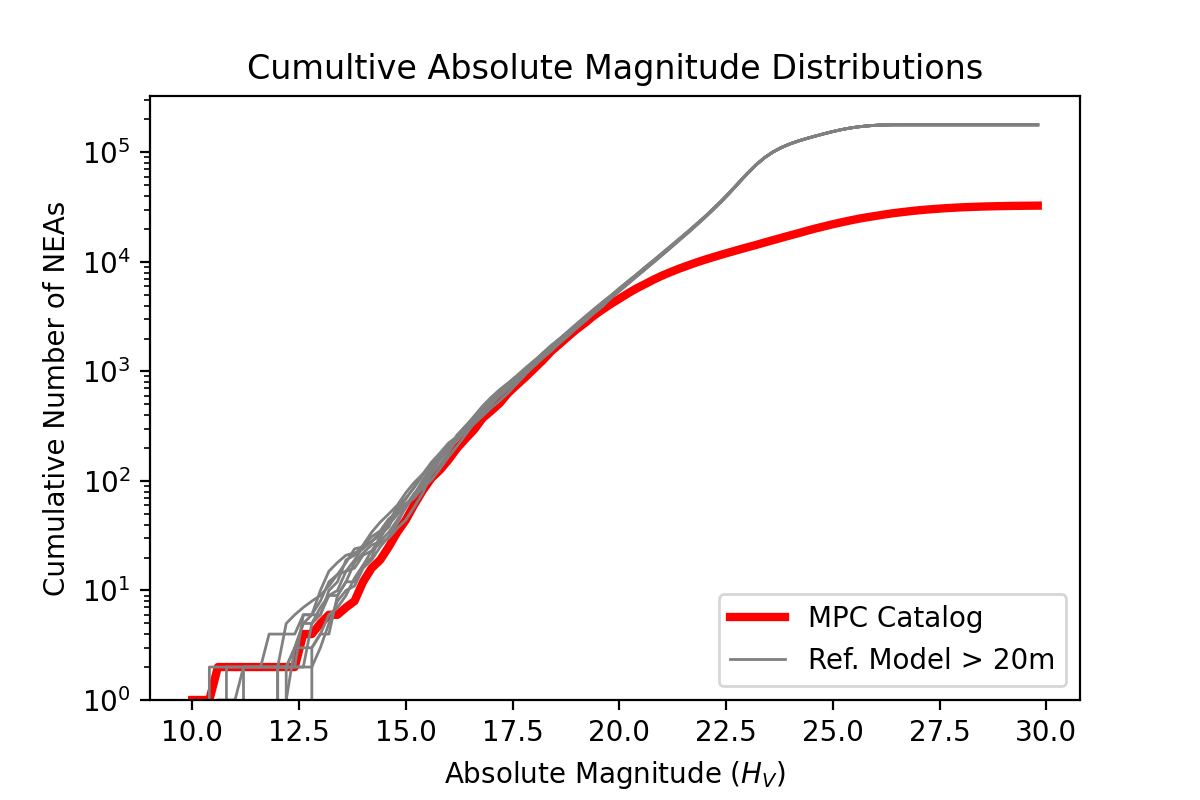}
\caption{The absolute magnitude distribution of ten instances of the NEOS reference model for NEAs (solid grey lines) and the known population from the MPC catalog (solid red line) as of April 2023.
\label{fig:hmag_dist}}
\end{figure}

The NEO Surveyor project team has built a simulation tool, the NEOS Survey Simulator \citep[NSS;][]{Mainzer.2015a,Grav.2016a,Mainzer.2023,Masiero.2023a}, to understand the performance of the mission in making progress towards its design and the George E. Brown, Jr. goal.  An integral part of evaluating the success of the NEO Surveyor mission is the use of the NEO Surveyor reference model, which contains a synthetic NEA population that the project uses as a ``yardstick" against which progress is measured. The NEA population of the reference model contains $\sim 25,000$ objects with diameters larger than 140 meter. A Kernel Density Estimator \cite[KDE;][]{Scott.1992a,scipy2020} method using the NEA known population with $H_V < 20$ mag serves as the input for the orbital elements, and the NEOWISE dataset \citep{Mainzer.2011e,Mainzer.2012b} is used as the basis for the model objects' physical properties (see \cite{Mainzer.2023} for a more in depth discussion of the NSS and the NEO Surveyor reference model). While the reference model is built using a size frequency and albedo distribution based on the NEOWISE results, Figure \ref{fig:hmag_dist} shows that the model is consistent with the absolute magnitude distribution found in the MPC catalog as of end of March of 2023. All the results described in this work are based on running at least 10 instances of the NEA reference model through our modelling in a Monte Carlo approach. 

%% Figure 5

%% ========================= SECTION ======================================
\section{NSS Known Object Model}
\label{sec:kom}

In order to understand the progress of the NEO Surveyor mission towards the George E. Brown, Jr. goal it is necessary to understand which of the synthetic objects in the reference model represent the objects that have already been found and cataloged by the MPC. The NSS uses a simple model to mimic the performance of the historical surveys over the last few decades. The model starts in 1970 and runs forward, tracking which objects in the reference model would be detected and when. For our analysis purposes we assume all discovery ends at the date of analysis, in this case end of 2022, in order to understand a worst-case scenario. We can also extend the model to predict which discoveries would be made up to the 2027 NEO Surveyor launch date for a best-estimate analysis as well. 

The NSS model for determining what portion of the NEAs in the  NEO Surveyor reference model are to be considered discovered at the start of the NEO Surveyor mission, hereafter called the Known Object Model (KOM), uses four parameters that change with time. The four parameters controlling the KOM are the limiting magnitude, the size of the field of regard in ecliptic longitude and latitude, and the chance of discovery if the object is found to be inside the field of regard and is brighter than the limiting magnitude. The model uses time steps of 30 days and computes the ecliptic position of each object in an instance of the reference model for each time step. The brightness of each object, V, is calculated using $V = H_V + 5 \log(\Delta *r) - 2.5 \log(F(G,\alpha))$, where $\Delta$ is the observer to object distance, $r$ is the heliocentric distance of the object, $\alpha$ is the phase angle of the object, and $F()$ and $G$ are the phase function and phase coefficient defined in \cite{Bowell.1989a}. The field of regard is centered on the opposition point as seen from the Earth and on the ecliptic plane, and has a half-width and half-height as shown in Table \ref{tab:model_parameters} (Columns 3 and 4, respectively). If the object is found to be in the field of regard and brighter than the limiting magnitude (see column 2 of Table \ref{tab:model_parameters}) for the time step considered, the object has a chance of being considered discovered equal to that given in column 5 in Table \ref{tab:model_parameters}. 

%% Figure 6
\begin{figure}[ht!]
\plotone{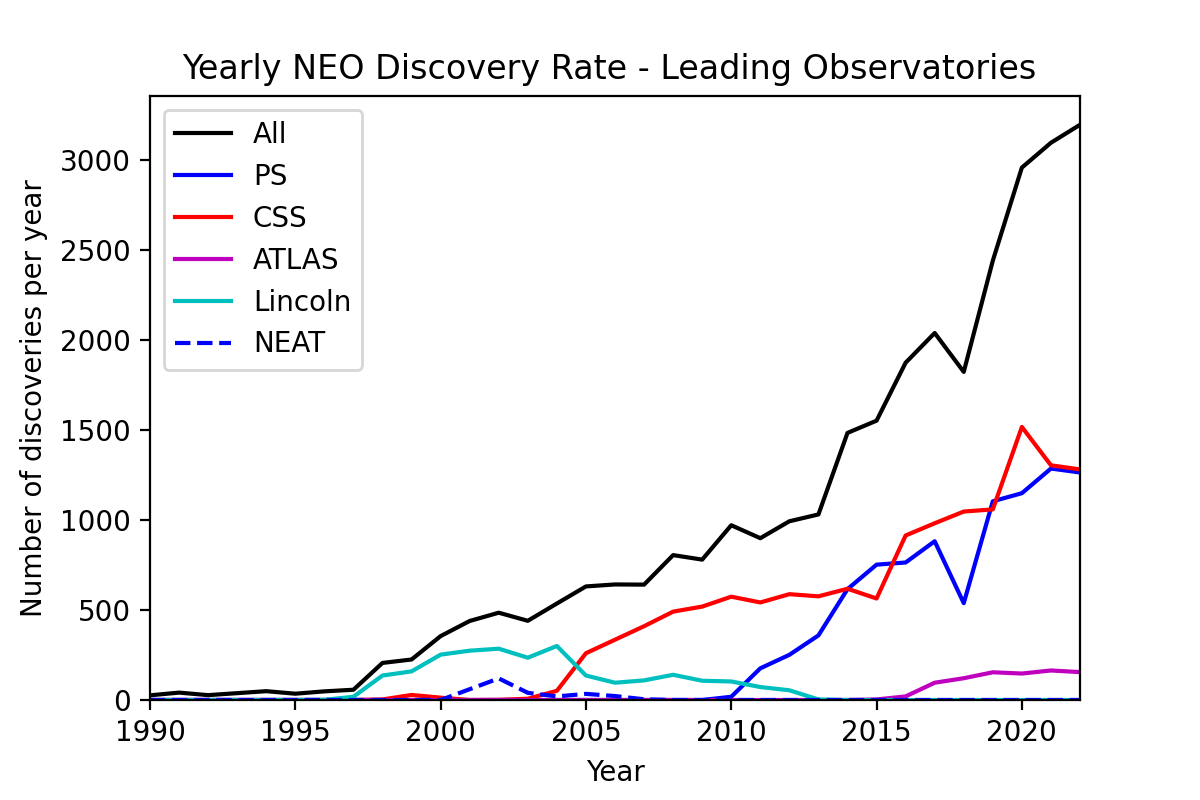}
\caption{The total discovery rate of NEAs per year is shown in solid black line, with the yearly discovery rate of NEAs  shown in different colored lines: Pan-STARRS (PS; MPC sites \#F51 and \#52) in solid blue, Catalina Sky Survey (CSS; MPC sites \#703 and \#G96) in solid red.
\label{fig:yearly_disc}}
\end{figure}

The values given in Table \ref{tab:model_parameters} are derived through a combination of methods. The first step is data analysis of the MPC catalog (downloaded on June 30, 2023) to determine which observatories dominate the various time periods over the last century. By the end of the 1970s only 80 NEAs were known, having been discovered by 22 different observatories, of which Palomar was the only site with more than a dozen discoveries \citep{vanHouten.1970a,vanHouten.1984a}. The next decade, from 1980-1989, saw a doubling of the number of known NEAs to 185, with 10 additional telescope recording discoveries. Palomar, MPC site \#675, dominated with 67 discoveries in this decade, with no other site achieving double digit NEA discoveries. Over the next six years, Palomar continues their work with $\sim 10$ new discoveries per year, but this time period sees the rise of the Spacewatch survey at Kitt Peak, MPC site \#691, which discovers 134 new NEAs over this time period \citep{Gehrels.1996a}. Spacewatch was quickly superceded by the dawn of the Lincoln Near-Earth Asteroid Research \citep[LINEAR;][]{Stokes.2000a,Stuart.2001a} program, MPC site \#704, in 1997. LINEAR was the first observatory to reach more than 100 discoveries in the following year. LINEAR continued dominating the discovery of NEAs for 7 years, discovering an average of 234 NEAs per year, until being superceded as the leading NEA discovery site by the Catalina Sky Survey (CSS), MPC sites \#703 and \#G96, in 2005 \citep{Larson.2007a,Zavodny.2008a,Granvik.2018}. CSS held the position as the leading yearly NEA discoverer for almost a decade, discovering an average of 477 new NEAs each year over this period. The next major change in the search for NEAs came in 2011, with the introduction of the first of the Pan-STARRS telescopes \citep[PS;][]{chambers2019panstarrs1,Denneau.2013a}, MPC site \#F51. By 2013 the field, led by the Catalina Sky Survey and Pan-STARRS, had increased the number of discoveries to more than a thousand NEAs per year, of which more than $90\%$ were discovered by those two dominating surveys. By the following year, Pan-STARRS had matched the Catalina Sky Surveys in NEA discoveries per year at 616 new NEAs for PS and 618 new NEAs for CSS that year. In 2016 the CSS improved the equipment at their MPC site \#G96, more than doubling the discoveries pear year at that site. This was followed just two years later in 2018 by the introduction of the second telescope, MPC site \#F52, by Pan-STARRS, which gave this project a modest boost of $\sim40\%$ new discoveries per year.

%% Figure 7
\begin{figure}[ht!]
\plotone{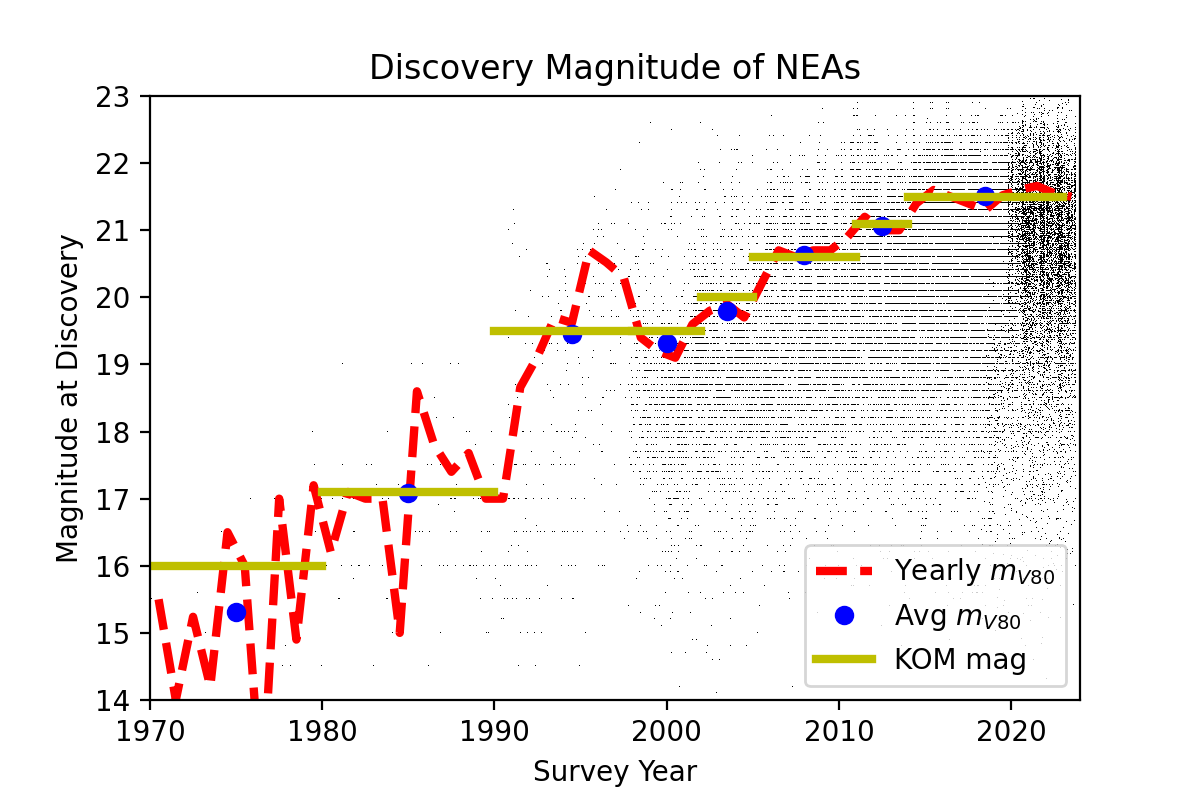}
\caption{The discovery epoch magnitude reported to the MPC for each NEA cataloged from 1970 to 2023. The yearly 80th-percentile value of the discovery magnitude, $m_V80$ is given as red dashed line, with the blue dots representing the average value of the $m_V80$ for each time period given in Table \ref{tab:model_parameters}. The yellow lines represent the value of the magnitude used in the KOM from Table \ref{tab:model_parameters}.
\label{fig:discmagNEAs}}
\end{figure}

For each of these time periods we use the discovery observations in the MPC catalog from the known population of asteroids in the inner solar system (excluding any objects discovered past the orbit of Jupiter) to determine a starting point for each of the four parameters (see Figure \ref{fig:discmagNEAs} for an example of the magnitude parameter). Once the starting point has been determined the KOM is run for a grid of values around these starting values, comparing the number of objects in the MPC catalog with the number of objects identified by the KOM as known for each time period and the absolute magnitudes of less than 17.75 magnitude, and less than each magnitude from 19 to, and including, 23. The grid steps start at 0.5 magnitudes for the limiting magnitude, with five degree steps for each field of regard parameter, and 5\% for the chance of discovery parameter. If none of the grid points yield a average of less than 10\% difference for each magnitude limit at the end of the each time period, the grid points are halved and the KOM is run for each of the new grid points. The resulting model parameter values represents the average value for that time period. We caution that the KOM is a simple first order model that uses average values over the stated time periods. Additional parameters such as rate of motion cuts or different latitude limits for each hemisphere are not considered as initial tests show that this added granularity does not significantly improve the model results.

\begin{figure}[ht!]
\plotone{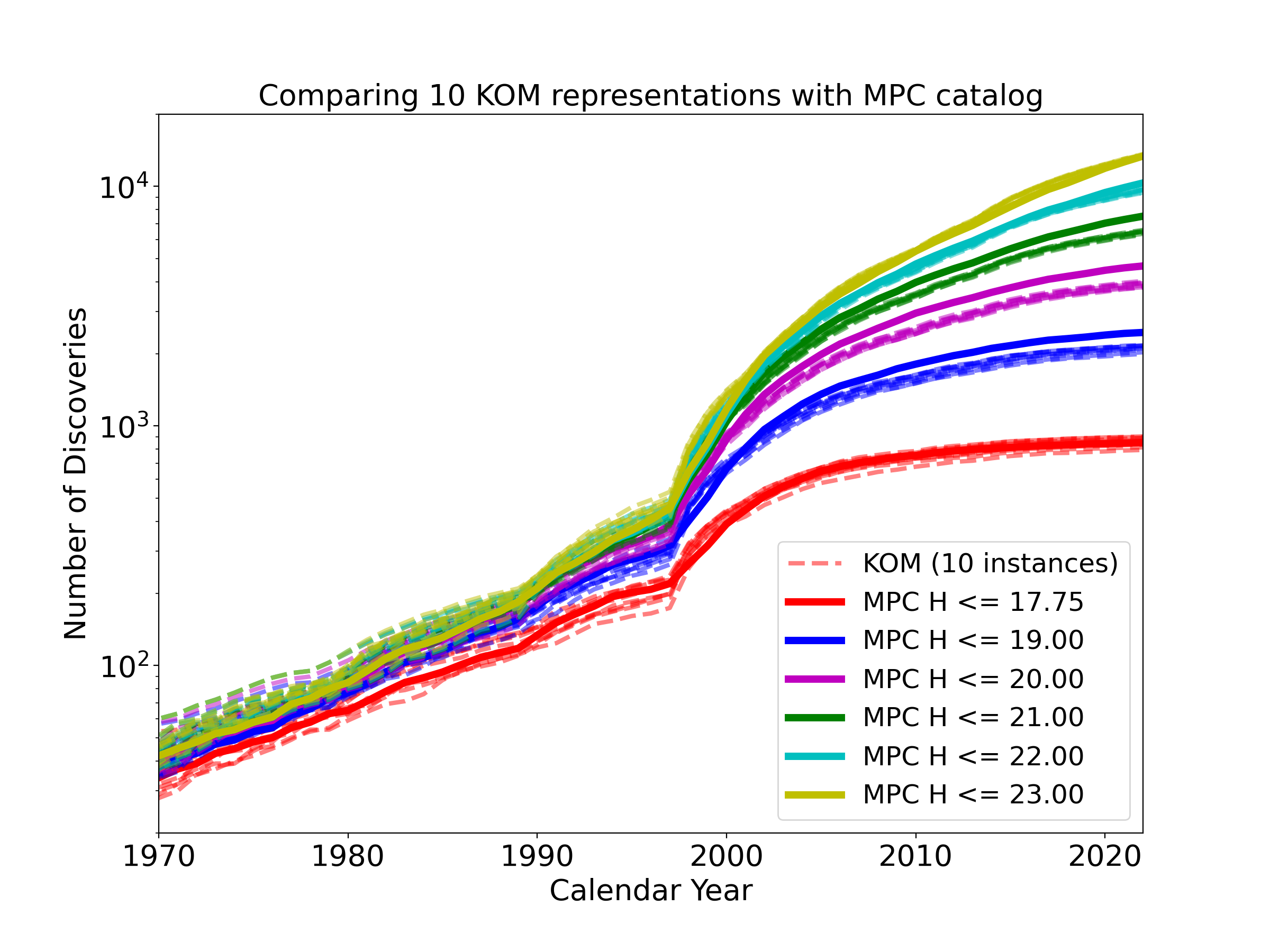}
\caption{The comparison of the KOM using the NSS reference model with the known NEA population from the Minor Planet Center at a range of absolute magnitude limits. The 10 random realizations of the NEA reference model were generated with a minimum effective diameter of 40m. 
\label{fig:kom_vs_mpc}}
\end{figure}

Figure \ref{fig:kom_vs_mpc} shows the discoveries of NEAs with $H_V$ limits ranging from $17.75$ magnitude (the traditional proxy of the 1 km objects) to 23 magnitude (the proxy of 140 m objects) from 1980 to 2022 from both the MPC catalog and the KOM.
The most prominent feature is the sharp increase in discoveries of NEAs in 1999, coinciding with the emergence of the Lincoln Near-Earth Asteroid Research (LINEAR) survey run by Lincoln Labs in New Mexico \citep{Stokes.2000a}. Another slight increase in happens in the mid-2000s with the introduction of the Catalina Sky Survey \citep{Larson.2007a}, operated by the University of Arizona, and then again in early 2010s by the introduction of the Pan-STARRS project, operated by the University of Hawaii \citep{chambers2019panstarrs1,Denneau.2013a}. 

\begin{table}[ht!]
    \centering
    \begin{tabular}{|l|c|cc|c|l|}
    \hline
    Years & Limiting & \multicolumn{2}{c|}{Field of Regard} & Chance of & Time Period \\
        & Magnitude & Longitude & Latitude & Discovery & \\
        &  (V mag)   & (degrees) & (degrees) & & \\
    \hline
     1930-1949 & 14.0 & 30 & 15  & 50\% & Historic Period I\\
     1950-1954 & 16.0 & 40 & 20 & 10\% & Historic Period II\\
     1955-1959 &      &     &     &  0\% & Historic Period II \\
     1960-1969 & 15.5 & 25 & 20 & 10\% & Historic Period III\\
     1970-1979 & 16.0 & 30 & 25 & 25\% & Historic Period IV\\
     1980-1989 & 17.1 & 30 & 20 & 30\% & Palomar Dominance\\
     1990-1997 & 19.5 & 30 & 25 & 10\% & Spacewatch Dominance \\
     1998-2001 & 19.5 & 40 & 30 & 85\% & LINEAR Start-Up\\
     2002-2004 & 20.0 & 45 & 30 & 90\% & LINEAR Dominance\\
     2005-2010 & 20.6 & 45 & 30 & 75\% & CSS Dominance \\
     2011-2013 & 21.1 & 45 & 30 & 75\% & CSS Dominance \& PS Startup\\
     2014-     & 21.5 & 45 & 30 & 90\% & PS \& CSS Joint Phase\\
     \hline
    \end{tabular}
    \caption{The parameters of the Known Object Model used to determine which NEAs in the model should be considered already discovered at the start of the NEO Surveyor mission.}
    \label{tab:model_parameters}
\end{table}

%% Figure 8

Applying this model to an instance of the reference model, one can then compare the objects that are considered discovered in the reference model to that of the known population from the MPC catalog. Figure \ref{fig:kom_vs_mpc} shows the number of discoveries of NEA for different limits of absolute magnitude, $H_V$, from $17.75$ to $23$ mag, which approximately spans the range from 1 km to 140 m. The number of discoveries reported to the MPC are given in by the solid lines, while 10 instances of the NEA reference model (using a minimum diameter of 40m) are plotted as dashed lines. Note that a model going down to 40m is needed to evaluate the discovery rate down to the 140m proxy of $H_V \sim 23$ mag since an object with effective diameter of 40m and an albedo of 0.5 would have an absolute magnitude of $H_V \sim 23.4$ mag. The KOM follows the major trends of the known population at all absolute magnitude cuts modeled. The final number of KOM objects with $H_V \leq 17.75$ mag identified as found at the end of 2022 were an average of $4.7\%$ more abundant than objects with $H_V < 17.75$ mag in the MPC catalog of known objects. A total of ten random realizations of the NEA reference model were created and processed through the KOM to assess the uncertainty (see Table \ref{tab:model_res}). For the next three absolute magnitude limits of $H_V \leq 19$ mag, $H_V \leq 20$ mag and $H_V \leq 21$ mag, the KOM underestimates the number of discovered objects by $12.1\%$, $13.1\%$ and $8.3\%$, respectively. For the $H_V < 22$ mag set the KOM is in excellent agreement with the corresponding set from the MPC catalog, only overestimating the number of known objects by $1.7\%$. The difference grows again for the $H_V \leq 23$ mag where the KOM over-estimates the number of known objects by $12.0\%$ on average over the 10 randomly generated realizations of the NEA reference model. 

\begin{table}[ht!]
    \centering
    \begin{tabular}{|r|ccc|ccc|ccc|}
        \hline
        Model & \multicolumn{3}{|c|}{$H_V < 17.75$ mag} & 
            \multicolumn{3}{|c|}{$H_V < 22$ mag} &
            \multicolumn{3}{|c|}{$H_V < 23$ mag} \\
         Year & MPC & Model & Diff (\%) & MPC & Model & Diff (\%) & MPC & Model & Diff (\%) \\
        \hline
            1949 & 16 & $15\pm5$ & $+7.3\pm33.4$ 
                 & 19 & $17\pm6$ & $+17.5\pm35.3$
                 & 19 & $20\pm6$ & $+17.5\pm35.3$ \\
            1959 & 23 & $25\pm7$ & $+9.1\pm28.3$
                 & 29 & $30\pm7$ & $+3.8\pm25.4$
                 & 29 & $30\pm7$ & $+4.8\pm24.8$ \\
            1969 & 33 & $32\pm8$ & $-2.7\pm25.6$ 
                 & 40 & $40\pm9$ & $+0.8\pm21.4$  
                 & 40 & $41\pm8$ & $+2.5\pm19.8$ \\
            1979 & 63 & $63\pm6$ & $+0.6\pm10.2$ 
                 & 80 & $84\pm8$ & $+4.5\pm10.1$ 
                 & 80 & $85\pm8$ & $+6.4\pm9.6$ \\
            1989 & 118 & $123\pm10$ & $+4.4\pm8.7$ 
                 & 185 & $187\pm13$ & $+1.1\pm7.3$ 
                 & 185 & $191\pm14$ & $+3.4\pm7.3$ \\
            1997 & 220 & $210\pm19$ & $-4.5\pm8.7$ 
                 & 440 & $443\pm25$ & $+0.6\pm5.7$
                 & 455 & $471\pm31$ & $+3.5\pm6.6$ \\
            2001 & 447 & $464\pm20$ & $+3.7\pm4.5$ 
                 & 1465 & $1442\pm29$ & $-1.5\pm2.0$ 
                 & 1556 & $1595\pm33$ & $+2.5\pm2.1$ \\
            2004 & 603 & $598\pm27$ & $-0.8\pm4.5$ 
                 & 2504 & $2423\pm55$ & $-3.2\pm2.2$ 
                 & 2704 & $2758\pm60$ & $+2.0\pm2.2$ \\
            2010 & 753 & $744\pm32$ & $-1.2\pm4.3$ 
                 & 4742 & $4482\pm60$ & $-5.5\pm1.3$ 
                 & 5372 & $5419\pm63$ & $+0.9\pm1.2$ \\
            2013 & 798 & $787\pm37$ & $-1.4\pm4.6$ 
                 & 5909 & $5701\pm68$ & $-3.5\pm1.1$
                 & 6876 & $7130\pm73$ & $+3.7\pm1.1$ \\
            2022 & 852 & $860\pm36$ & $+1.0\pm4.2$ 
                 & 10334 & $9600\pm120$ & $-7.1\pm1.2$
                 & 13378 & $13458\pm111$ & $+0.6\pm0.8$ \\
        \hline
    \end{tabular}
    \caption{Comparing the average number of objects found using 10 instances of the reference model with the number of known objects at the end of each time period in Table \ref{tab:model_parameters}.}
    \label{tab:model_res}
\end{table}

The differences between the sample of synthetic objects identified by the Known Object Model as known at the end of 2022 and those existing in the MPC catalog as seen in Table \ref{tab:model_res} are not surprising. Not only is the KOM a simplistic model, but the reference model makes a number of assumptions at the smaller sizes due to the lack of information at sizes below $140$ m. Specialty surveys, such as those conducted at low ecliptic solar elongation and NEOWISE \citep{Mainzer.2011e,Mainzer.2012b}, account for about $\sim 10\%$ of the discovered objects. These surveys generally have shallower limiting magnitudes or cover less area than the dominant opposition surveys. Since the KOM does not model these surveys, it is expected that it falls short in identifying objects in the NEA population, especially for $H_V \sim 18.5-20.5$ mag, which is the absolute magnitude range over which most of the specialty surveys have found objects that are not observable by the opposition surveys. For example, NEOWISE has discovered 393 NEAs by the end of 2022. Of these, 170 have $H_V \leq 20$ mag which alone accounts for almost a quarter of the average difference of 730 objects when the KOM applied to 10 instances of the reference models are compared to the MPC catalog. For the $H_V \leq 21$ mag range, NEOWISE has discovered 274 NEAs, which accounts for $\sim30\%$ of the difference between the KOM and the MPC catalog. On the flip side at $H_V \leq 22$ mag the 328 NEA discoveries by NEOWISE would represent an increase of $\sim 3\%$ to the difference between the model and catalog. To put these number in better perspective we can examine the sample of $H_V \leq 23$ mag, the new proxy for 140 meter NEAs. At this size range \citet{Mainzer.2011e,Mainzer.2012b} estimates that the uncertainty in our models are $\pm3000$ for NEAs larger than 100 meter. According to \citet{Harris.2021a} the completeness level at $H_V \leq 23$ mag is $\sim 30\%$, which means that our model assumptions introduce an error of $\pm 1000$ NEAs, which is significantly higher than the influence of surveys like NEOWISE not being included in the KOM. 

The difference between the KOM model and the known catalog of NEAs described above may also be partially driven by the assumptions used to generate the NEA reference model. The over-abundance of objects found in the KOM model for NEA with $H_V$ is consistent with our reference model having an overabundance of objects with $H_V \sim 23$ mag. \citep{Mainzer.2011e,Mainzer.2012b} estimated that there are $20,500 \pm 3,000$ NEAs larger than 100 meter. The NASA NEO Science Definition Report \citep[SDT;][]{Stokes.2017a} increased this to $\sim 25,000$ NEOs larger than 140m while acknowledging that this new estimate was larger than the previous result. This was attributed to using a single size-frequency distribution slope between $70m < D < 1.5km$, but they believed that the differences reflect the uncertainties that exists in the current size-frequency estimates. Thus, by selecting a reference model with 25,000 NEAs larger than 140 meter in line with the SDT report \citep{Mainzer.2023}, the difference between the KOM and the MPC catalog of 1602 objects represents only $\sim 35\%$ of this uncertainty, which is on the same order as the completeness for $H \leq 23$ mag derived by \citet{Harris.2021a}. \citet{Heinze.2021a} also found that for NEOs smaller than $H_V \sim 22.5$ mag are more common than expected using extrapolation of the size-frequency distribution at larger sizes. In conclusion, it is clear that our understanding of the size-frequency distributions at smaller than $H_V \sim 22.5$ mag is not well understood, which is exactly the range where a survey like NEO Surveyor will be discovering most efficient. It is thus not unreasonable to conclude that the difference between the model presented in this paper and the MPC catalog at $H \leq 23$ mag are primarily driven by an over-abundance of these objects in our reference model.  

Some other reasons for the differences between the number of objects detected by the KOM and the number of objects cataloged by MPC have been considered. As seen in Figure \ref{fig:subpop_fraction}, the fractions based on the known population remain relatively stable over the absolute magnitude range covered in this analysis, and the relative fraction of objects as a function of diameter is kept constant during the construction of the NEA reference model. However, the distribution among the sub-populations shown in Figure \ref{fig:subpop_fraction} is based on the raw observed data and may contain observational biases that are not well understood and therefore not incorporated into the KOM. Other similar assumptions, such as the albedo distribution being the same across all diameter bins for each sub-population, may also not hold, but these assumptions remain the best knowledge we currently have about the NEA population. Furthermore, the reference model makes a number of assumptions at the smaller sizes due to the lack of information at sizes below $140$m. It is possible that a break in the SFD exists somewhere around $\sim 100$ m, although the location and magnitude of this break remain uncertain \citep{Harris.2015a,Harris.2021a,Granvik.2018}. Such a break would change the relative number of NEAs in the different absolute magnitude limit samples, with objects of 100m having absolute magnitudes of $H_V \sim 21-25$ mag. It is in part due to the uncertainty of the validity of these assumptions that the future generation of surveys, such as NEO Surveyor, are of such importance. These future surveys are key in testing our current assumptions and knowledge about the NEA population and will provide new and improved NEA reference models that will help us better understand the danger the NEA population poses.

%% Figure 9
\begin{figure}[ht!]
\plotone{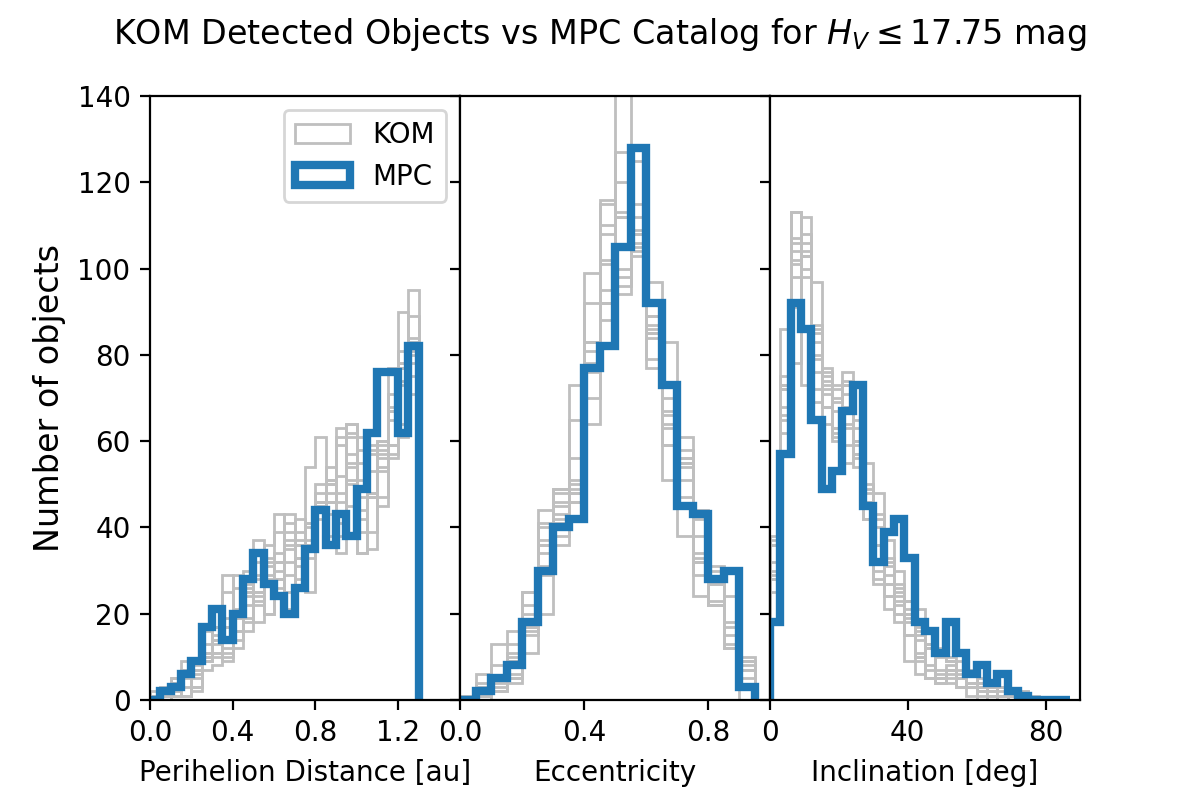}
\caption{The comparison of the Known Object Model using the reference model with the known NEA population from the Minor Planet Center for objects with $H_V \le 17.75$ mag (the proxy for objects larger than $\sim 1$km). The orbital elements of perihelion distance, eccentricity and inclination are shown from left to right, respectively. The KOM applied to 10 random realizations of the NEA reference model is shown in gray, while the MPC distribution is shown in blue.
\label{fig:model_vs_known_orbele_h1775}}
\end{figure}

%% Figure 10
\begin{figure}[ht!]
\plotone{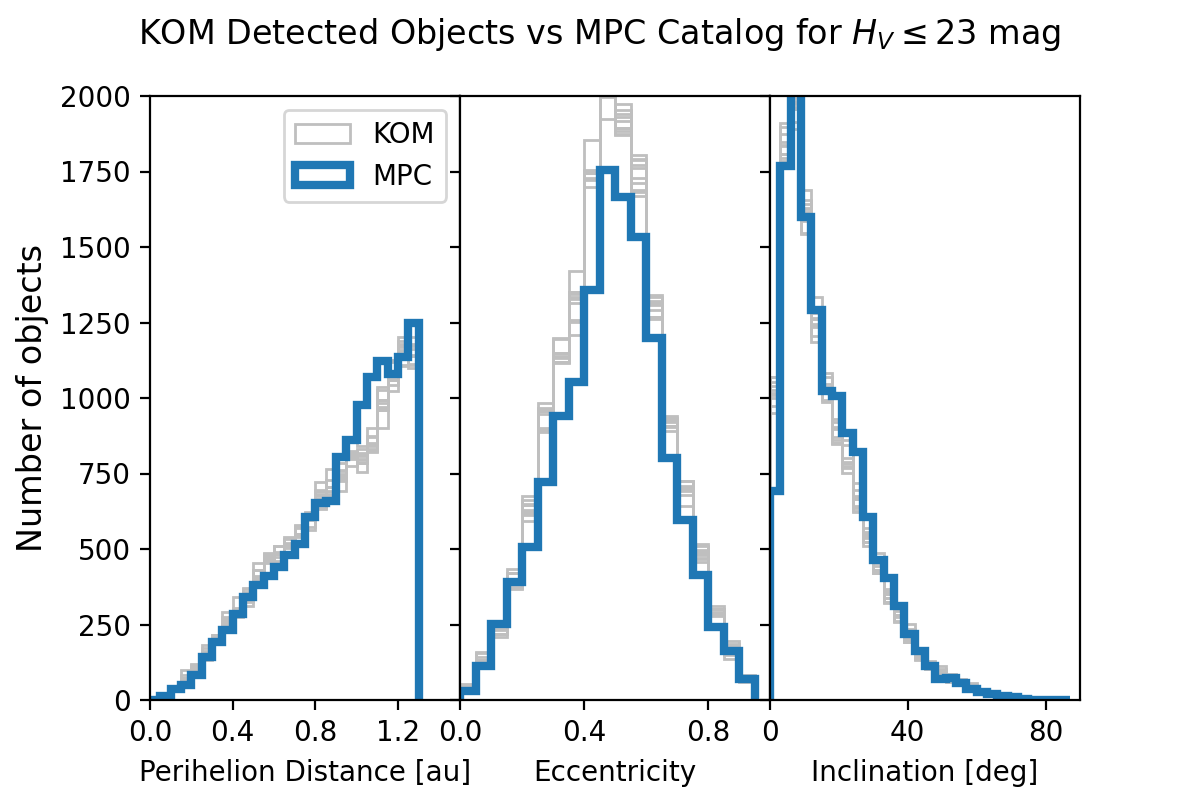}
\caption{The comparison of the Known Object Model using the NEA reference model with the known NEA population from the Minor Planet Center for objects with $H_V \le 23$ mag (the proxy of objects larger than $\sim 140$ m). The orbital elements of perihelion distance, eccentricity and inclination are shown from left to right, respectively. The KOM applied to ten random realizations of the NEA reference model is shown in gray, while the MPC distribution is shown in blue.
\label{fig:model_vs_known_orbele_h23}}
\end{figure}

While the KOM provides a reasonable estimate the discovery rates of the MPC catalog over the last few decades for a range of absolute magnitude bins, it is also important to make sure that the model is identifying all types of objects regardless of orbital elements and physical properties. Figure \ref{fig:model_vs_known_orbele_h1775} and \ref{fig:model_vs_known_orbele_h23} compare the orbital elements of the synthetic population the KOM identified as found compared to the orbital elements of the known objects in the MPC catalog for two absolute magnitude limits. Both figures show agreement between the distributions in semi-major axis, eccentricity, and inclination. One noticeable difference in the $H \leq 23$ mag sample is a slight underestimation of found Amor objects in the Known Object Model, which indicates that the Amor population may be slightly underestimated in the NEO Surveyor reference model. 

\begin{figure}[ht!]
\plotone{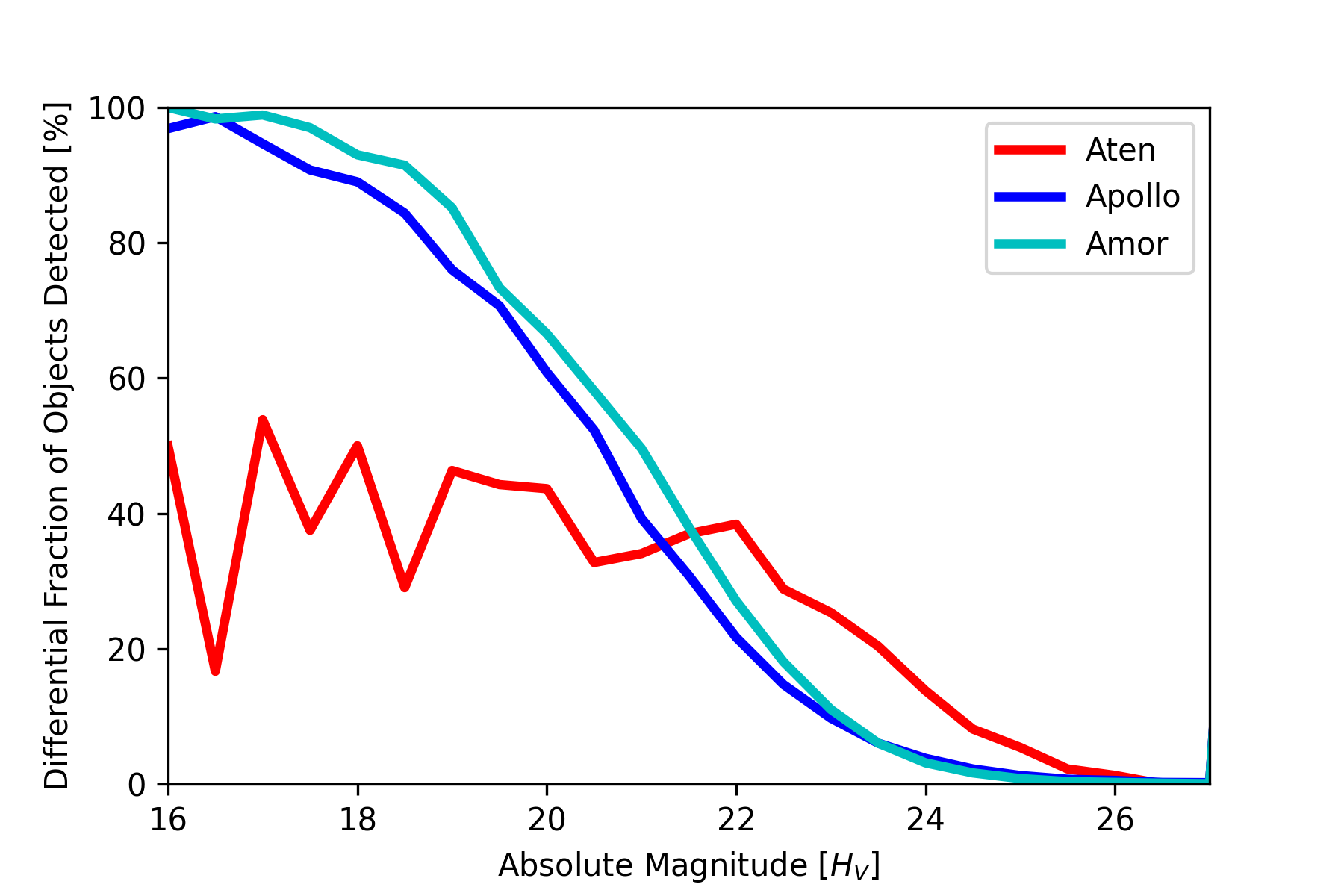}
\caption{The fraction of objects observed by the KOM as a function of absolute magnitude $H_V$ in bins of $\pm0.25$mag around the absolute magnitude. The Amor sub-population is favored by almost 10\% for $H_V < 23$ mag over the Apollo sub-population. For objects with $H_V > 23$ mag, there is little difference in the biases of these two sub-populations. The Atens have significantly higher chances of being detected by the KOM for $H_V > 22$ mag. 
\label{fig:obs_biases}}
\end{figure}

We can also look at the observational biases that exist in a pure opposition survey, which is what the KOM represents. The lack of any Atira asteroids detected in the KOM is one of the most clear biases, but there are also interesting biases among the other sub-populations. When studying the fraction of objects detected as a function of $H_V$ (see Figure \ref{fig:obs_biases}), it is seen that for the larger objects, $H_V \sim 16$ mag, the Amor sub-population is favored by almost $3\%$ higher completeness over objects in the Apollo sub-population. This difference increases to almost $10\%$ higher completeness at $H_V \sim 21$ mag. At $H_V > 23$ mag the difference in chance of detection drops to close to zero between these two populations. For the Aten population, the observational biases are significant at larger sizes, with only $\sim 40\%$ of objects detected in the $18 < H_V < 22$ mag range. At fainter magnitudes, $H_V > 23$ mag, opposition surveys are more likely to detect Atens, relative to both Apollos and Amors. These biases are a combination of the observational geometry and the on-sky location where the survey is operating, primarily centered on opposition.

%% Figure 11

%% Figure 12
\begin{figure}[ht!]
\plotone{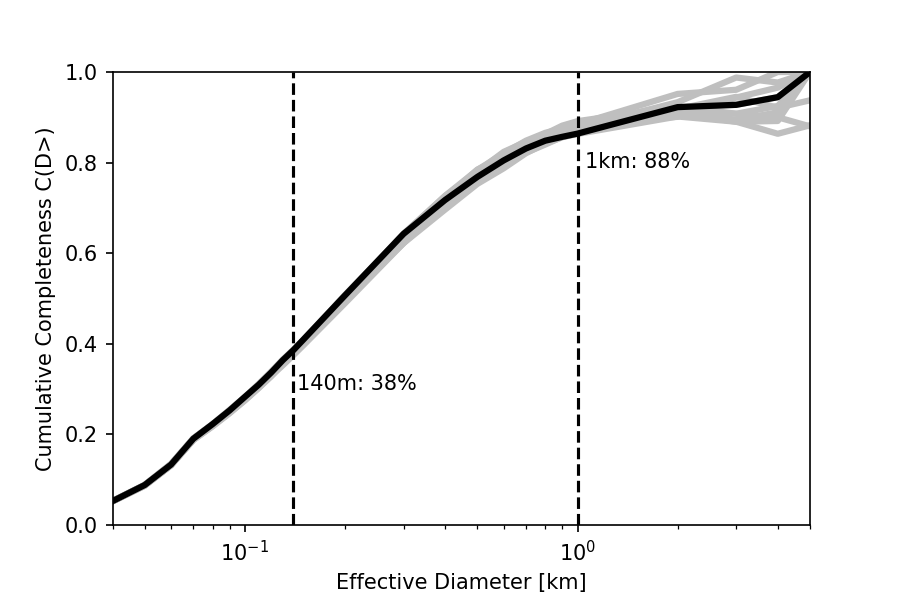}
\caption{Cumulative completeness of the diameter, $C(D>)$, at end of 2022 as a result of the Known Object Model computed for ten random realizations of the NEA reference model with minimum size set to 40 m. The average cumulative completeness is shown as the solid line, showing that the KOM estimates an average completeness of $88\%$ for NEAs larger than 1km and $38\%$ for NEAs larger than 140 m. The discrepancy between our completeness estimate at 1 km and the estimates of $90+\%$ completeness from \citet{Mainzer.2011a,Granvik.2018} are due to the fact that the KOM does not include specialty surveys (such as the low solar elongation surveys, NEOWISE, etc.). NEOWISE has, for example, contributed at least 55 NEA discoveries with diameters larger than 1 km, which would account for $\sim6\%$ of completeness for this size range. 
\label{fig:Dcompleteness_annotated}}
\end{figure}

%% Figure 13
\begin{figure}[ht!]
\plotone{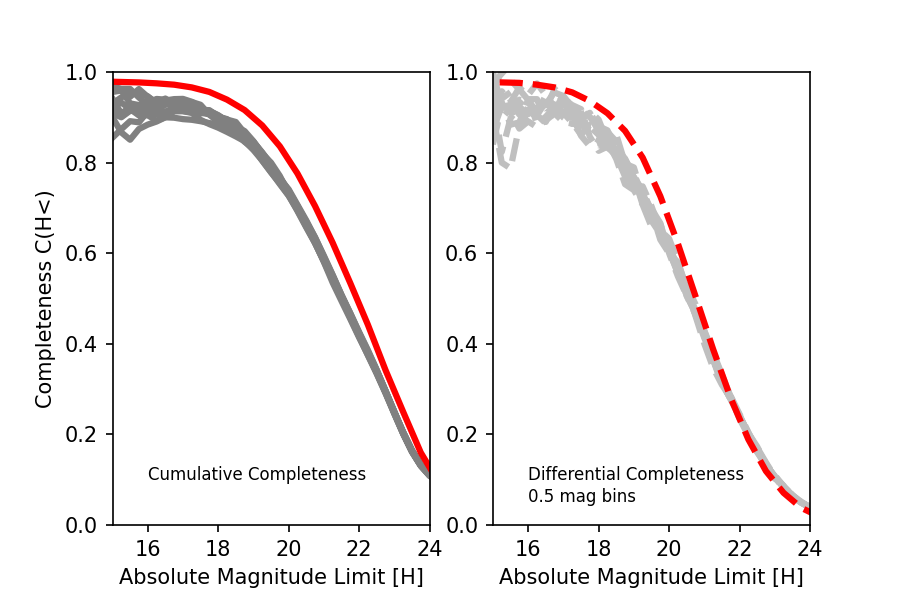}
\caption{Cumulative and differential completeness of the absolute magnitude, $C(H_V<)$, at end of 2022 as a result of the Known Object Model for ten instances of the reference model with minimum size set to 40m. The differential completeness is shown as the light grey solid lines, while cumulative completeness is shown as the dark gray lines. The cumulative and differential completeness as a function of absolute magnitude as derived by \cite{Harris.2021a} are shown as the solid red line and the dashed red line, respectively. The missing completeness in the KOM at $H_V < 20$ mag are due to the KOM not modeling the low solar elongation surveys and NEOWISE.  
\label{fig:kom_vs_harris}}
\end{figure}

Assuming that the Known Object Model is a reasonable estimate of the known objects currently cataloged by the MPC, we examine the completeness of the the current catalog. The cumulative completeness as a result of the KOM for objects with diameters larger than a specific diameter limit is shown in Figure \ref{fig:Dcompleteness_annotated}. For the larger objects with $D \geq 1$ km, the KOM finds a completeness of $87\%$, which is slightly below the $90\%$ completeness found by \citet{Mainzer.2011a,Granvik.2018}. This is mainly due to the lack of low solar-angle surveys in the KOM, with NEOWISE alone having contributed 351 new discoveries not accounted for in the KOM. Of these, at least 55 are larger than 1km \citep{Nugent.2015a,Nugent.2016a,Masiero.2017a,Masiero.2020a,Masiero.2021a}. On average our 10 random realizations of the NEA reference model have $997\pm39$ NEAs larger than 1 km, with the KOM model yielding a completeness on average of $88\pm1\%$ for this size regime. Thus the additional discoveries provided by NEOWISE represent a significant portion of the difference between the completeness derived by the KOM and that found by \cite{Mainzer.2011a} and \cite{Granvik.2018}.

For the objects larger than 140 m, the KOM returns a completeness of $\sim 38.3\pm0.3\%$ at the end of 2022, when applied to 10 instances of the reference model. This is consistent with the completeness derived by \cite{Harris.2021a}, which found $44\%$ completeness for $H_V < 22.25$ mag, $34\%$ completeness for $H_V < 22.75$ mag, and $25\%$ completeness for $H_V < 23.25$ mag. \cite{Harris.2021a} rely mainly on optical observations and thus derived completeness estimates, both cumulative and differential, as functions of absolute magnitude. Figure \ref{fig:kom_vs_harris} shows their results compared to the completeness modeled by the KOM across different absolute magnitude limits. It shows that KOM underestimates the completeness down to $H_V \sim 21$ mag, which is consistent with the low-solar elongation surveys, such as NEOWISE, not being modelled by KOM. At $H_V \leq 23$ mag, our proxy for effective diameter of 140 m, the KOM and \cite{Harris.2021a} results are in reasonable agreement. Note that the flattening of the cumulative distribution at $H_V > 24$ mag is due to limit in $H_V$ imposed by limiting the reference model used to 40m or larger.

%% ========================= SECTION ======================================
\section{Known Object Model objects and NEO Surveyor}
\label{sec:NEOSandKOM}

One of the questions faced by the NEO Surveyor mission is understanding which synthetic objects identified as known by the Known Object Model would and would not be seen by the observatory over its nominal mission. Objects that are both observed in the optical by the ground based surveys (such as those modeled as detected by KOM) and detected in the thermal by NEO Surveyor during its nominal 5-year survey are of particular interest as both their diameters and albedo can be determined. When applying the NSS with a nominal 5 year survey planning model to the reference model for NEAs described in \citet{Mainzer.2023}, we find that for NEAs larger than 140m, $\sim 77\%$ of the $\sim9750$ objects that the KOM identified as currently known will also be detected and tracked by NEO Surveyor during its nominal survey. Thus, almost one third of the NEAs larger than 140m will have both optical and thermal observations collected at the end of the NEO Surveyor 5-year mission, providing diameter and albedo for an order of magnitude more NEAs with diameters larger than 140m than are available today \citep{Mainzer.2019a,Masiero.2021a}. Of course, there will also be numerous smaller NEAs with both optical and thermal measurements that will also significantly contribute to the understanding of the NEA populations' physical properties.

\begin{figure}[ht!]
\plotone{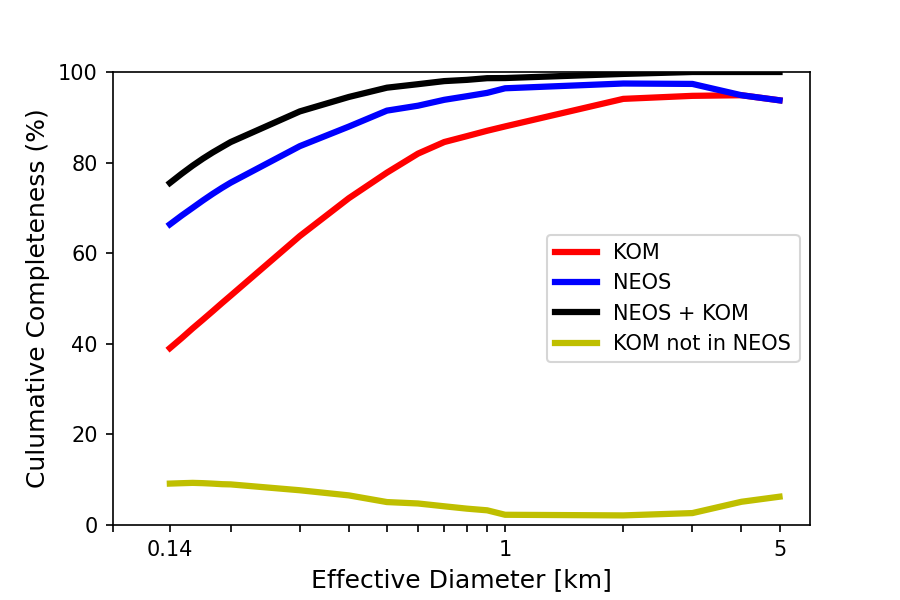}
\caption{The cumulative completeness of the effective diameter for NEAs in the reference model for the Known Object Model (red solid line), the NEO Surveyor mission (blue solid line) and the combined result of the KOM and the NEO Surveyor Mission (black solid line). Also shown is the fraction of objects seen by KOM that NEO Surveyor is not likely to detect in its 5-year nominal mission (yellow solid line). 
\label{fig:kom_vs_neos_cumcomp}}
\end{figure}

Another set of objects of interest is the objects that are already known that will not be detected by NEO Surveyor during its nominal mission. While neither diameter or albedo can be determined for these objects, knowledge of their orbital parameters means that they will remain part of the catalog of known objects, contributing to the final PHA and NEA completeness reached at the end of the NEO Surveyor survey. The remaining objects, those that were identified as already known by the KOM, but are not likely to be detected by the NEO Surveyor, contribute $8-9\%$ to the total completeness of the catalog at the end of the NEO Surveyor nominal mission. Figure \ref{fig:kom_vs_neos_cumcomp} shows that the KOM-estimated survey completeness for NEAs larger than 140 m is $\sim38\%$ as of the end of 2022. Also shown is the completeness of a single random realization of the reference model for NEAs larger than 140 m, simulating the performance of the NEO Surveyor's nominal 5 year survey. Without any prior knowledge from ground-based surveys, the cumulative completeness from NEO Surveyor's 5 year nominal survey is $\sim66\%$ for this size regime. When combining the KOM and the NEO Surveyor results, the final catalog of known objects would be $\sim76\%$ complete. For the PHAs, the completeness is predicted to be $82\%$ complete \citep{Mainzer.2023} at the end of the NEO Surveyor 5-year mission. As expected, NEO Surveyor and the Known Object Model combine to find nearly all NEAs larger than 1km. 

%% Figure 14

%% Figure 15
\begin{figure}[ht!]
\plotone{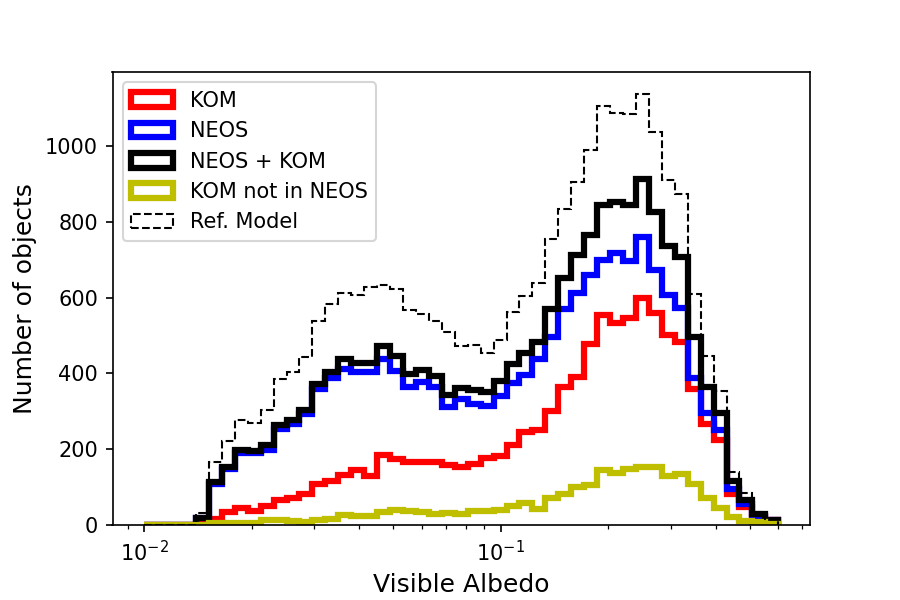}
\caption{The visible albedo distribution of the objects that are larger than 140m from an instance of the reference model identified by the Known Object Model (red histogram) as currently cataloged. The objects identified as cataloged by NEO Surveyor are given as a blue histogram, with the combined set of objects cataloged by bith KOM and NEOS shown as a black histogram. The NEAs larger than 140m that are identified as cataloged by the KOM and not seen by the NEO Surveyor in a nominal 5 year survey are shown as the yellow histogram.
\label{fig:kom_vs_neos_albedo}}
\end{figure}

When looking at the albedo distributions, the observational biases towards the higher albedo NEAs are clearly seen in the objects identified as cataloged by the Known Object Model (see Figure \ref{fig:kom_vs_neos_albedo}). This feature was pointed out in \cite{Mainzer.2011e,Mainzer.2012b}. As with the NEOWISE results, NEO Surveyor is almost free of biases in albedo and detects objects of all albedos nearly equally well.

%% Figure 16
\begin{figure}[ht]
\plotone{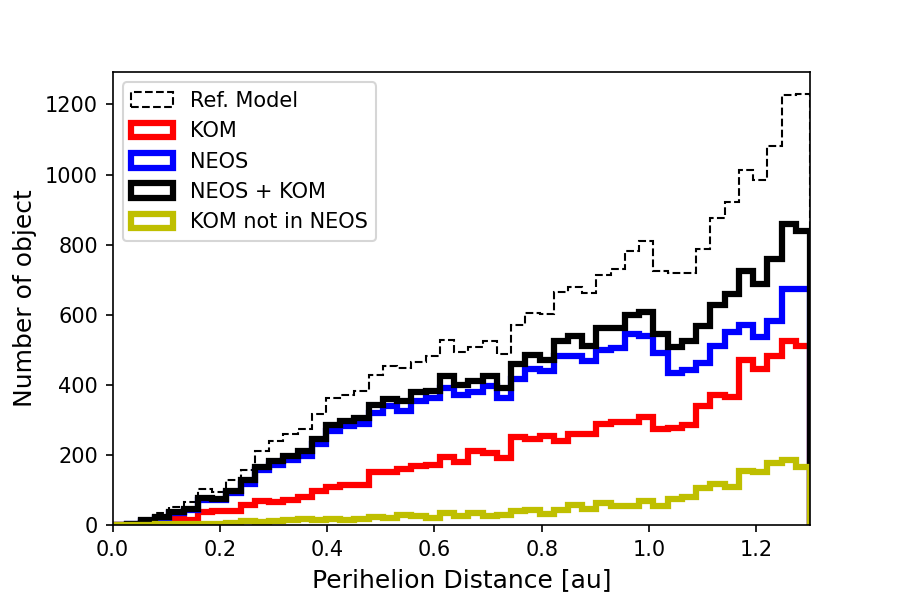}
\caption{The distribution of perihelion distance for the objects cataloged by KOM up until the end of 2022 (red histogram), NEO Surveyor 5-year nominal mission (blue histogram) and the two combined (black solid histogram). The distribution of objects identified as cataloged by KOM that are not observed by NEO Surveyor are shown as a yellow histogram. This shows that majority, at $\sim 59\%$, of known objects missed by the NEO Surveyor are Amors, where the current surveys are fairly efficient due to the favorable geometry. 
\label{fig:kom_vs_neos_peridist}}
\end{figure}

%% Figure 17
\begin{figure}[ht!]
\plotone{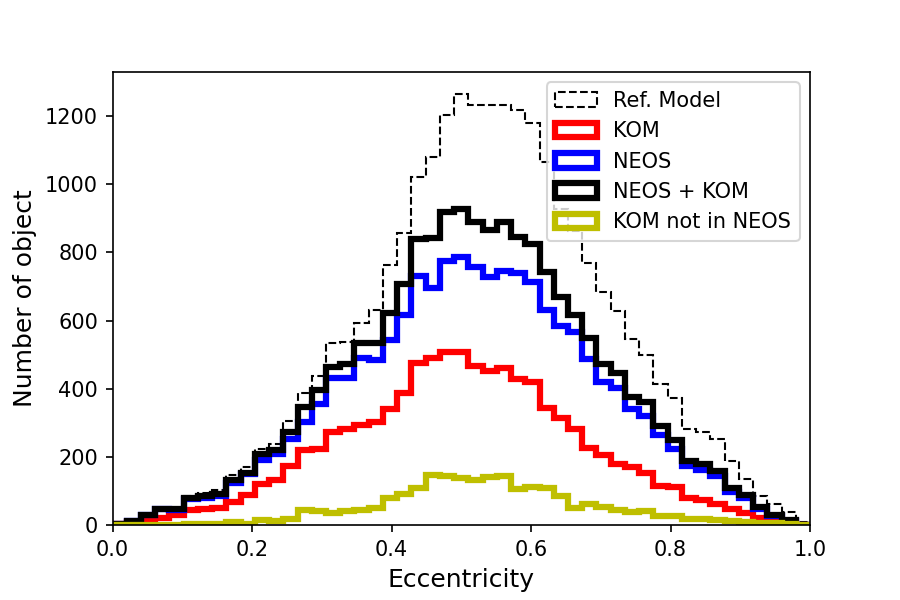}
\caption{The eccentricity distribution of the objects cataloged by KOM up until the end of 2022 (red histogram), NEO Surveyor 5-year nominal mission (blue histogram) and the two combined (black solid histogram). The distribution on objects identified as cataloged by KOM that are not observed by NEO Surveyor are shown as a yellow histogram. As expected the NEO Surveyor is very efficient at detecting and tracking the low eccentricity objects, similarly to the current surveys.
\label{fig:kom_vs_neos_ecc}}
\end{figure}

\begin{figure}[ht!]
\plotone{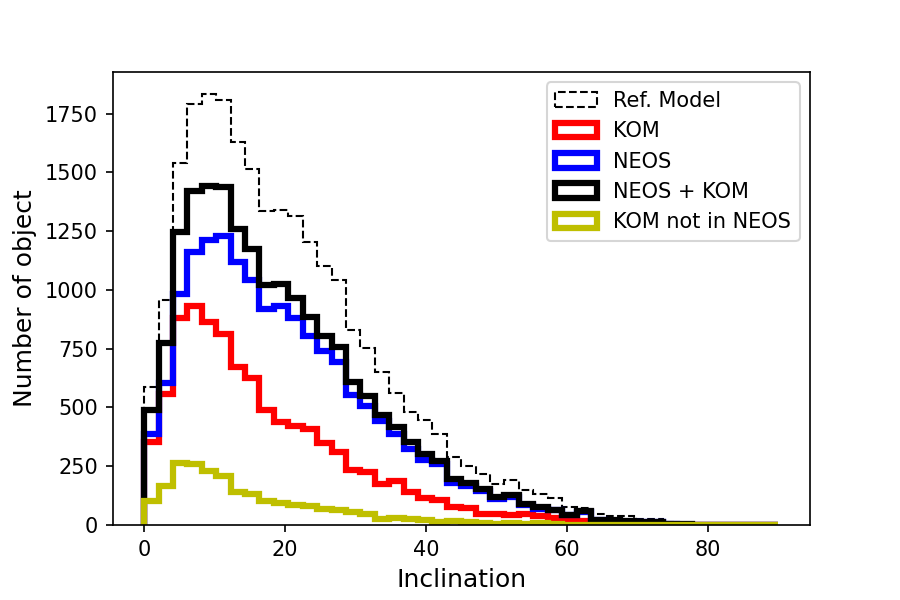}
\caption{The inclination distribution of the objects cataloged by KOM up until the end of 2022 (red histogram), NEO Surveyor (blue histogram) and the two combined (black solid histogram). The distribution of objects identified as cataloged by KOM that are not observed by NEO Surveyor are shown as a yellow histogram. 
NEO Surveyor is found to be slightly more efficient at detecting and tracking the higher inclination NEOS ($i > \sim15$ degrees) than the current surveys. 
\label{fig:kom_vs_neos_incl}}
\end{figure}

Examining the orbital elements of the objects identified by the KOM and NEO Surveyor (see Figure \ref{fig:kom_vs_neos_peridist}, \ref{fig:kom_vs_neos_ecc}, and \ref{fig:kom_vs_neos_incl}) show that the ground based surveys, according to the KOM, are slightly more efficient at detecting Amors ($\sim42\%$ completeness at end of 2022) compared to the Apollos ($\sim38\%$ completeness) and Atens ($\sim35\%$ completeness) for objects larger than 140m. For the same size regime NEO Surveyor is more efficient at detecting Atiras, Atens and Apollos than Amors, with NEOs detecting $\sim88\%$ of Atiras, $\sim95\%$ of Atens, $\sim71\%$ of Apollos, and $\sim57\%$ of Amors during its 5-year mission. The synergy between the currently known object catalog and the objects detected and tracked by NEO Surveyor becomes apparent when looking at the $\sim9\%$ of NEAs in the reference model larger than 140m that are identified as cataloged and not seen by NEO Surveyor (see Figure \ref{fig:kom_vs_neos_peridist}). A majority, $\sim59\%$, of this set of objects are Amors, followed by $\sim40\%$ Apollos and less than one percent that are Atens.

When looking at the eccentricity distributions (see Figure \ref{fig:kom_vs_neos_ecc}), the combination of the currently cataloged objects and the objects discovered by NEO Surveyor account for almost all objects with eccentricity less than $0.4$. The objects with higher eccentricity spend more time further away from the Earth's orbit and need to be closer to the perihelion point in their orbits to be detected and tracked. This can be remedied by extending the NEO Surveyor survey duration, giving these objects more additional time to approach their perihelia when passing through the NEO Surveyor field-of-regard. The inclination distributions (see Figure \ref{fig:kom_vs_neos_incl}) show similar trends, with nearly all low inclination objects being cataloged by a combination of the KOM and the NEO Surveyor. A vast majority of the objects that are not cataloged have inclinations of 15 degrees or more. 

%% ========================= SECTION ======================================
\section{Conclusions}
\label{sec:conclusions}
In this paper we have shown that using a simple model, called the Known Object Model, we can provide a reasonable estimate of which NEAs in the reference population model created for the NEO Surveyor mission are currently known. These objects represent the combined efforts of mainly ground-based surveys such as Lincoln Near-Earth Asteroid Research \citep[LINEAR;][]{Stokes.2000a,Stokes.2002a}, Catalina Sky Survey \citep{Larsen.2007a} and the Pan-STARRS project \citep{Wainscoat.2010a,chambers2019panstarrs1}. The simple model approximately recreates the NEA discovery rate recorded by the MPC catalog over a wide range of absolute magnitudes, $H_V \sim 17 - 23$ mag. 

When applying the KOM to a set of randomly generated realizations of the NEO Surveyor reference model, it is estimated that the catalog completeness of NEAs larger than 140 m at the end of 2022 stands at $\sim 38\%$, which is consistent with the results of \cite{Harris.2021a} when using $H_V < 23$ mag as a proxy for this population. It is further found that $\sim 77\%$ of the objects larger than 140 m cataloged by the KOM are identified as also being detected by NEO Surveyor in a 5 year nominal survey. The remaining set of objects larger than 140m cataloged by the KOM, and not detected by NEO Surveyor, represents $\sim 9\%$ of the total number of NEAs larger than 140 m in the NEOS reference model. These two sets of objects can be combined with the objects cataloged by NEO Surveyor to derive a modeled cataloged completeness of $\sim 76\%$ for NEAs larger than 140 m and $\sim 82\%$ for PHAs larger than 140 m at the end the NEO Surveyor 5 year baseline mission \citep{Mainzer.2023}.

\section{{Acknowledgments}}
\begin{acknowledgments}
This publication makes use of data products from the NEO Surveyor, which is a joint project of the University of Arizona, and the Jet Propulsion Laboratory/California Institute of Technology, funded by the National Aeronautics and Space Administration. 

This publication also makes use of data products from NEOWISE, which is a project of the Jet Propulsion Laboratory/California Institute of Technology, funded by the National Aeronautics and Space Administration. 

This publication also makes use of data products from the Wide-field Infrared Survey Explorer, which is a joint project of the University of California, Los Angeles, and the Jet Propulsion Laboratory/ California Institute of Technology, funded by the National Aeronautics and Space Administration.
\end{acknowledgments}


\vspace{5mm}
%%\facilities{}


\software{NumPy \citep{numpy2020a},
            SciPy \citep{scipy2020},
            Astropy \citep{astropy.2022a},
            Jupyter \citep{jupyter2016}
          }




\bibliography{}
%%\bibliography{references}
\bibliographystyle{aasjournal}



\end{document}